\documentclass[twocolumn,english,aps,superscriptaddress,prl,floatfix,nofootinbib]{revtex4-2}
\usepackage[latin9]{inputenc}
\usepackage{color}
\usepackage{babel}
\usepackage{amsmath}
\usepackage{amssymb}
\usepackage{mathtools}
\usepackage{graphicx}

\newcommand{\fsso}{Fe$_4$Si$_2$Sn$_7$O$_{16}$}

\graphicspath{{./}{./figs/}}

\begin{document}
\title{Partial magnetic order in kagome spin ice}

\author{Eric C. Andrade}
\affiliation{Instituto de F\'isica, Universidade de S\~ao Paulo, C.P. 66318,
05315-970, S\~ao Paulo, SP, Brazil}
\author{Matthias Vojta}
\affiliation{Institut f\"ur Theoretische Physik and W\"urzburg-Dresden Cluster
of Excellence ct.qmat, Technische Universit\"at Dresden, 01062 Dresden,
Germany}
\begin{abstract}
Motivated by the observation of partial magnetic order in kagome-based magnets, we study the classical kagome Ising antiferromagnet, known as kagome spin ice, including further-neighbor interactions at zero and finite temperature. While the nearest-neighbor model displays an extensive ground-state degeneracy, various symmetry-breaking states can appear upon including additional couplings. Among these, peculiar partially ordered states have been proposed. We present results from large-scale Monte-Carlo simulations, establishing that such partial order is stabilized by third-neighbor couplings along the hexagon diagonals. We show that these states arise due to the emergence of one-dimensional chains in an entirely frustrated environment,  and that they are stable with respect to further couplings. We discuss their finite-temperature properties in detail, highlight the magnetic states' peculiarities depending on the diagonal couplings' sign, and suggest observing these states using thermodynamic probes.
\end{abstract}
\date{\today}
\maketitle


Highly frustrated magnets constitute a prime territory in the quest for novel phases of matter. Those include non-trivial forms of order and disorder, such as quantum spin liquids, spin nematics, as well as skyrmions and other topological magnetic textures \cite{lacroix11,castelnovo12,fert17,vojta18,fernandes19,sachdev23}.
A fascinating class of states is that with partial magnetic order, where a fraction of the magnetic moments display symmetry-breaking long-range order while another fraction remains fluctuating to low temperatures. Such partially ordered (or, equivalently, partially disordered) states have been reported for a few compounds, among them the heavy-fermion metals CePdAl \cite{oyamada08} and UNi$_4$B \citep{movshovich99}, as well as the insulating magnets Sr$_2$YRuO$_6$ \cite{granado13},  Gd$_2$Ti$_2$O$_7$ \cite{javanparast15},   LiZn$_2$Mo$_3$O$_8$ \cite{sheckelton12},   HoAgGe \cite{zhao20}, and  {\fsso} \cite{ling17,dengre21}.

Partially ordered states have also appeared in various theoretical models, including Ising models \cite{meka77,blan84, takagi93, wills02,moto12}, classical vector-spin models at finite $T$ \cite{diep97,javanparast15}, quantum Heisenberg models \cite{gonzalez19,seifert19}, and Kondo models with partial screening \cite{moto10}. In all cases, the critical ingredient is that the network of the ordered moments does not exert any mean field on the remaining disordered moments. The latter fluctuating magnetic moments effectively remove a fraction of the frustrated bonds while retaining entropy, thus lowering the free energy.  However, in such a situation, order by disorder \cite{villain80, shender82,henley89} would still be possible,  and the precise mechanism preventing a complete ordering is specific for each case. Notably, conclusive links between theory and experiment are rare \cite{javanparast15},  and further well-understood examples are paramount to building a general scenario.

Kagome ice \cite{wills02}, referring to spins on the two-dimensional kagome lattice with infinite local Ising anisotropy, belongs to the class of frustrated Ising models \cite{zhao20}. In its simplest version, it features a macroscopic ground-state degeneracy and a variety of order-by-disorder phases, resulting in a non-trivial temperature evolution of the magnetic states \cite{moller09,chern11,chern12,colbois22}.
A partially ordered phase has been proposed early on to occur in a model supplemented by second-neighbor interactions \cite{wills02}, and this proposal has been discussed vis-a-vis experimental data \cite{dengre21}. In this paper, we revisit the issue of partial order in kagome ice \cite{zhao20}.  Utilizing large-scale Monte-Carlo (MC) simulations, we show the previous proposal \cite{wills02} to be incorrect. At the same time, we identify a different parameter regime where partial order is indeed stabilized at low $T$ -- this is facilitated by including a third-neighbor coupling along the diagonals of the hexagons. The spins connected via this extra coupling form one-dimensional chains coupled by ice rules. Such exquisite arrangement leads to intense frustration and dimensional reduction because one out of three chains experiences a vanishing mean field and is entropically disordered at any finite $T$. The resulting model for ferromagnetic diagonal coupling displays two thermal phase transitions of Kosterlitz-Thouless (KT) type, with partial order emerging below the lower transition. We also investigate this model's antiferromagnetic cousin, finding a single transition and partial order at low $T$. We suggest the latter model being of relevance to {\fsso}.

\begin{figure}[!bt]
\includegraphics[width=1\columnwidth]{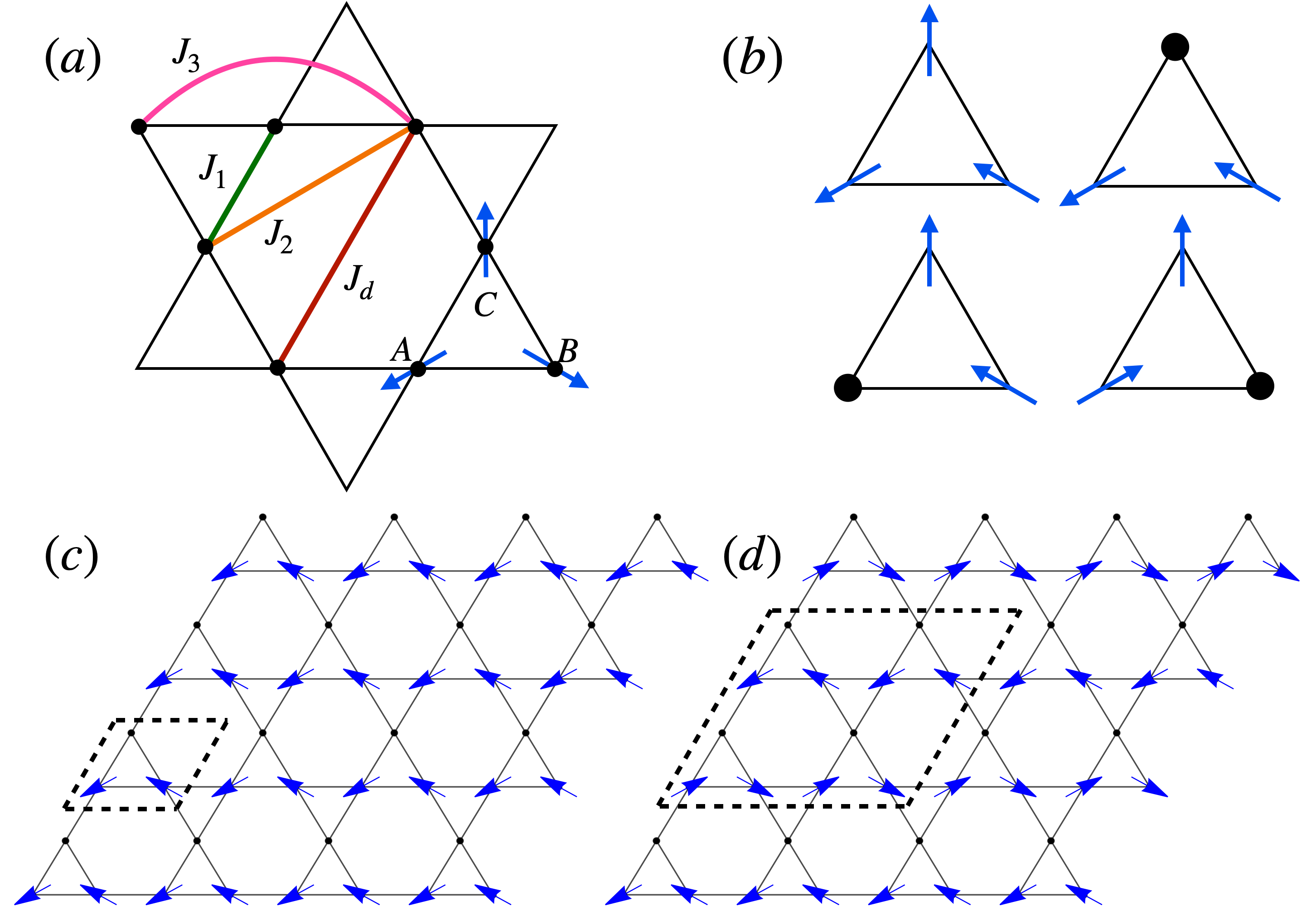}
\caption{\label{fig:states}
(a) Kagome lattice and its three sublattices, dubbed $A$,  $B$, and $C$. For each sublattice, we define the local axes $\hat{\boldsymbol{e}}_{i}$ as shown by the arrows. The exchange couplings are $J_1$, $J_2$,$J_3$, and $J_d$ between first-neighbor, second-neighbor,  third-neighbor across a site,  and diagonal spins,  respectively.
(b) Local spin configuration satisfying the ice rules, alongside three local configurations for the partially disordered state, all shown in the global frame.
(c,d) Partially disordered states,  in the global frame,  as favored by (c) $J_d>0$ and (d) $J_d<0$. The black dots indicate the disordered sites, while the dashed lines show the magnetic unit cell.
}
\end{figure}

\paragraph{Model.---}
The kagome ice model \cite{wills02} describes spins $\boldsymbol{S}_{i}=\sigma_{i}\hat{\boldsymbol{e}}_{i}$ on the vertices of the kagome lattice in the limit of infinite Ising anisotropy. The local axes $\hat{\boldsymbol{e}}_{i}$ connect the centers of adjacent triangles and point out of the upward triangles, and the $\sigma_{i}=\pm1$ encode the Ising degree of freedom,  see Fig.~\ref{fig:states}(a). In terms of the Ising degrees of freedom,  the Hamiltonian reads
\begin{align}
\begin{split}
\mathcal{H} & =
 J_{1}\sum_{\langle i,j\rangle}\sigma_{i}\sigma_{j}
+J_{2}\sum_{\langle\langle i,j\rangle\rangle}\sigma_{i}\sigma_{j}
-J_{d}\sum_{\langle\langle\langle i,j\rangle\rangle\rangle}\sigma_{i}\sigma_{j}\\&
-J_{3}\sum_{\langle\langle\langle i,j\rangle\rangle\rangle}\sigma_{i}\sigma_{j}
\end{split}
\label{eq:h}
\end{align}
where $J_{1,2,3,d}$ denotes first-neighbor, second-neighbor,  third-neighbor across a site,  and third-neighbor diagonal couplings,  respectively,  as in Fig.~\ref{fig:states}(a). The signs in Eq.~\eqref{eq:h} account for the relative orientation of the local Ising axes, such that $J_{1,2,3,d}>0$ corresponds to ferromagnetic exchange couplings in the global frame. A dominant $J_1$ imposes ice rules at low energy such that we have two spins in and one out or vice versa for each triangle, Fig.~\ref{fig:states}(b). We also note that, in applied fields, the saturated magnetization can be parallel to any of the three local axes, corresponding to a $Z_6$ symmetry.



\paragraph{Overview of parameter regimes.---}
Before describing our new results for the model in Eq.~\eqref{eq:h}, we quickly summarize those from the literature \cite{takagi93,wills02,moller09,chern11,chern12,colbois22}. For ferromagnetic $J_1>0$ and $J_{2,3,d}=0$, Eq.~\eqref{eq:h} displays an extensive manifold of degenerate ground states.  For $J_{2}<0$ and $J_{3,d}=0$, this macroscopic degeneracy is completely lifted, resulting in an ordered $\sqrt{3}\times\sqrt{3}$ state \cite{chern11,chern12}.  For $J_2>0$, Wills \textit{et al.} \cite{wills02} proposed a partially ordered phase to exist,  Fig.~\ref{fig:states}(c); our results below indicate that this proposal is incorrect.
In contrast, when all couplings are antiferromagnetic, the model displays macroscopically degenerate ground-state phases \citep{colbois22}, hence a finite residual entropy survives for $J_{2} > 0$ and  $J_{3,d} < 0$ in extended regions of the phase diagram,  showing that further couplings do not always select ordered ground states out of the disordered spin-ice manifold \cite{colbois22}.

To stabilize partial order as in Figs.~\ref{fig:states}(c) and (d), it appears plausible to introduce a diagonal coupling $J_d$ while setting $J_{2,3}=0$. This is easiest seen in the limit $\left| J_d \right| \gg J_1$: For $J_1=0$, the system decomposes into three Ising chains, which remain disordered at any finite $T$. A finite coupling $J_1$ locally imposes the ice rules and couples the chains but leaves a degeneracy. We then expect the partially ordered states in Figs.~\ref{fig:states}(c) and (d) to emerge for $J_1 > 0$, $J_d  \neq 0$, and $J_{2,3}=0$, by combining the effects of the spin-ice rules and the strong thermal fluctuations in spin chains. This model has not been studied extensively; this is the main target of this work.


\begin{figure}[!tb]
\includegraphics[width=1\columnwidth]{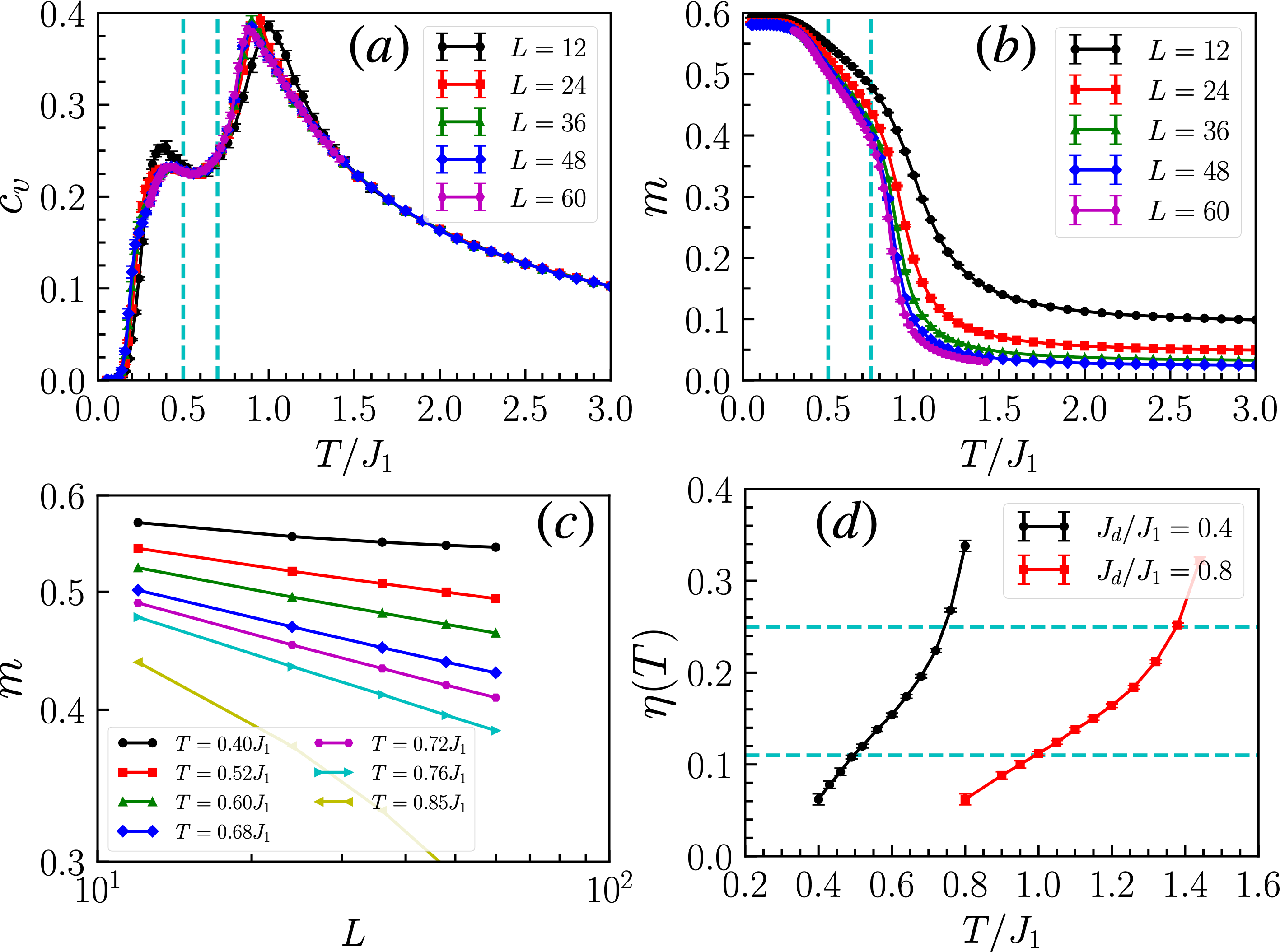}
\caption{
MC results for the kagome ice model with ferromagnetic $J_{d}/J_{1}=0.4$.
(a) Specific heat $C(T)$ for different linear system sizes $L$; the vertical dashed lines show the extent of the critical region (see text).
(b,c) Magnetization for the up-triangles in the global frame plotted as (b) $m(T)$ and (c) $m(L)$.
(d) Magnetization exponent $\eta(T)$ from $m\sim L^{-\eta/2}$, characterizing the critical region. Here we also show data for $J_{d}/J_{1}=0.8$.
}
\label{fig:cv_jdm}
\end{figure}

\paragraph{MC simulations.---}
We study Eq.~\eqref{eq:h} using classical MC simulations on lattices of linear size $L$, with three sites per unit cell,  Fig.~\ref{fig:states}(a), and periodic boundary conditions. The total number of sites is $N=3L^{2}$, and we consider $L=12$ up to $L=72$. We perform equilibrium MC simulations using single-site Metropolis updates combined with the parallel tempering method \cite{newman99}. Due to its highly frustrated nature, this system may not reach equilibrium at low $T$ sometimes, even for moderate system sizes. This, however, does not prevent us from extracting a general scenario for the thermodynamic behavior of Eq.~\eqref{eq:h}.

To characterize the putative partially ordered phases, we compute the magnetization in the global frame for each up-triangle inside the magnetic unit cell.  An average over all up-triangles in a given MC configuration returns a two-dimensional vector $\boldsymbol{m}=\left(m_x,m_y \right)$.  We define $m=|\langle \boldsymbol{m}\rangle|$ where the brackets $\langle\ldots\rangle$ denote the MC average. According to Fig.~\ref{fig:states}(b), $m \to 1/\sqrt{3}$ in the partially ordered phases as $T\to0$. It is useful to compute the histogram of the magnetization $P(m_{x},m_{y})$, which is expected to display peaks at $(m\cos\theta, m\sin\theta)$ with $\theta=\arctan\left(m_{y}/m_{x}\right)=n\pi/3$, $n=0,1,\ldots,5$ for $T \to 0$, different from a fully ordered state \cite{suppl}. We also compute the static spin structure factor, differentiating between the two partially ordered states, as they have different magnetic unit cells.
Finally we calculate the Edwards-Anderson order parameter, $q=\langle \sum_{i}\sigma_{i}^{\left(1\right)}\sigma_{i}^{\left(2\right)}\rangle /N$ where $(1,2)$ are replica indices. $q$ reflects the long-time autocorrelation function of a spin in an MC simulation \cite{lee07}. Therefore, $q=0$ in the paramagnetic phase while $|q|\to 1$ in a state with all spins frozen. In the partially ordered states, we expect $|q|\to4/9$ instead \cite{suppl}.


\paragraph{Partially ordered ferromagnet.---}
As motivated above, we focus on a model with positive $J_1$ and $J_d$, while $J_{2,3}=0$, to stabilize the partially ordered stripe configuration in Fig.~\ref{fig:states}(c). Thermodynamic data for $J_d/J_1=0.4$ are shown in Fig.~\ref{fig:cv_jdm}. The specific heat shows two peaks, with a weak dependence on system size, which we identify as phase transitions at $T_{c1}$ and $T_{c2}$, Fig.~\ref{fig:cv_jdm}(a). Both peak locations scale with $J_d$ and tend to merge as $J_{d}$ increases further. Also, for small $J_d$, there is a bump in the specific heat at $T/J_1 \approx 1.8$,  signaling the onset of ice rules \cite{wills02,suppl}.
The magnetization approaches $1/\sqrt{3}$ as $T\to0$, strongly hinting that we enter the partially ordered phase with $1/3$ of the spins fluctuating, Fig.~\ref{fig:cv_jdm}(b). Following Refs.~\onlinecite{takagi93,chern11,chern12}, we interpret the results as follows: Upon cooling from high $T$, we enter a critical intermediate phase with power-law correlations characterized by $m \sim L^{-\eta/2}$. Indeed, power-law fits to the magnetization data in Fig.~\ref{fig:cv_jdm}(c) represent the data well and yield the exponents shown in Fig.~\ref{fig:cv_jdm}(d). The critical phase expected for the six-state clock model is characterized by $\eta\left(T_{c1}\right)=1/4$ and $\eta\left(T_{c2}\right)=1/9$ \cite{jose1977,price12,price13}, which enables us to determine $T_{c1,2}$ more accurately. The KT character of the two transitions is also compatible with the weak dependence of $C(T)$ on system size. We finally enter the partially ordered phase for $T<T_{c2}$. Integrating the specific heat, we detect no extensive zero-point entropy, indicating that the residual fluctuations are correlated and the ground state is unique (up to discrete symmetries). This is consistent with the proposed partially ordered state whose entropy density vanishes as $T\to 0$ because the spin correlation length along the chains diverges, $\xi_{\rm{chain}} \gg L$.

\begin{figure}[!tb]
\includegraphics[width=1\columnwidth]{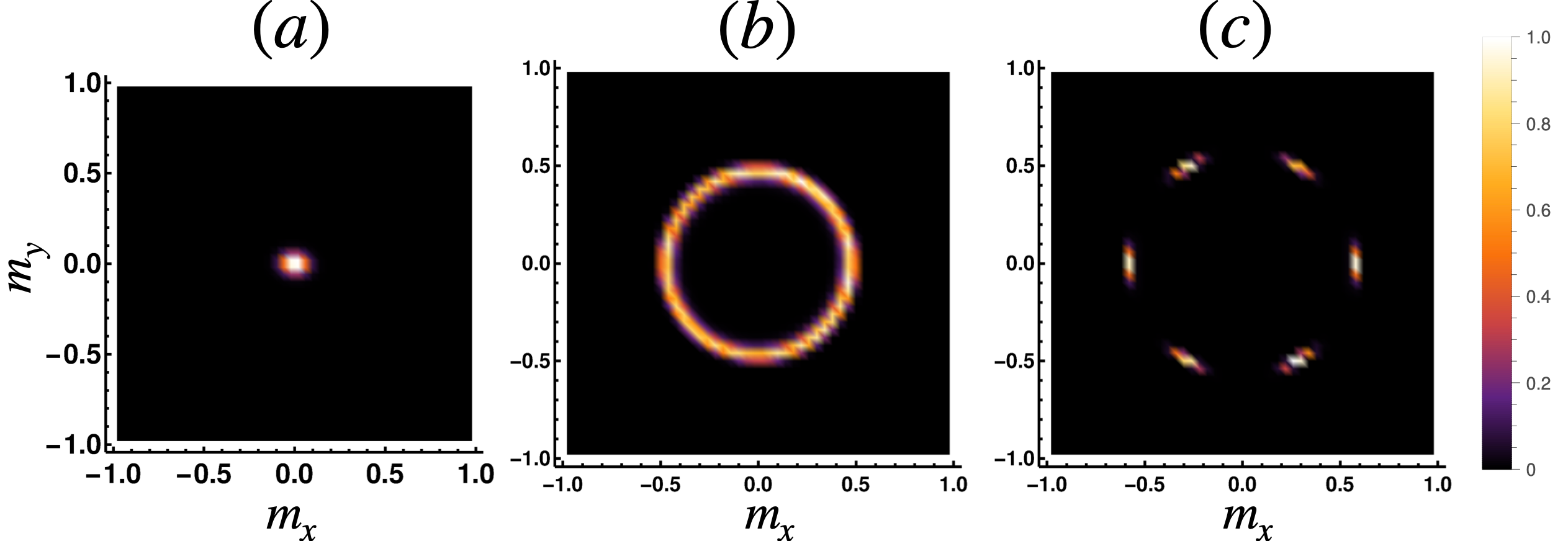}
\caption{\label{fig:P_mx_my_jdm}
Magnetization distribution $P(m_{x},m_{y})$ obtained for $L=48$ and $J_{d}/J_1=0.4$.
(a) $T/J_{1}=1.2$, (b) $T/J_{1}=0.6$, and (c) $T/J_{1}=0.3$.
}
\end{figure}


The histogram of the global magnetization $P\left(m_{x},m_{y}\right)$ is shown in Fig.~\ref{fig:P_mx_my_jdm}. In the high-temperature paramagnet, it is featureless and centered at $m=0$. The intermediate critical phase displays a ring-shaped structure, i.e., emergent U(1) symmetry \cite{lou07}.  For low $T$, it has the six-peak structure expected for the partially ordered phase, reflecting the emergent clock anisotropy of the model. Finally, the Edwards-Anderson order parameter reaches the expected value of $4/9$ at low $T$ \cite{suppl}.

\begin{figure}[!bt]
\includegraphics[width=1\columnwidth]{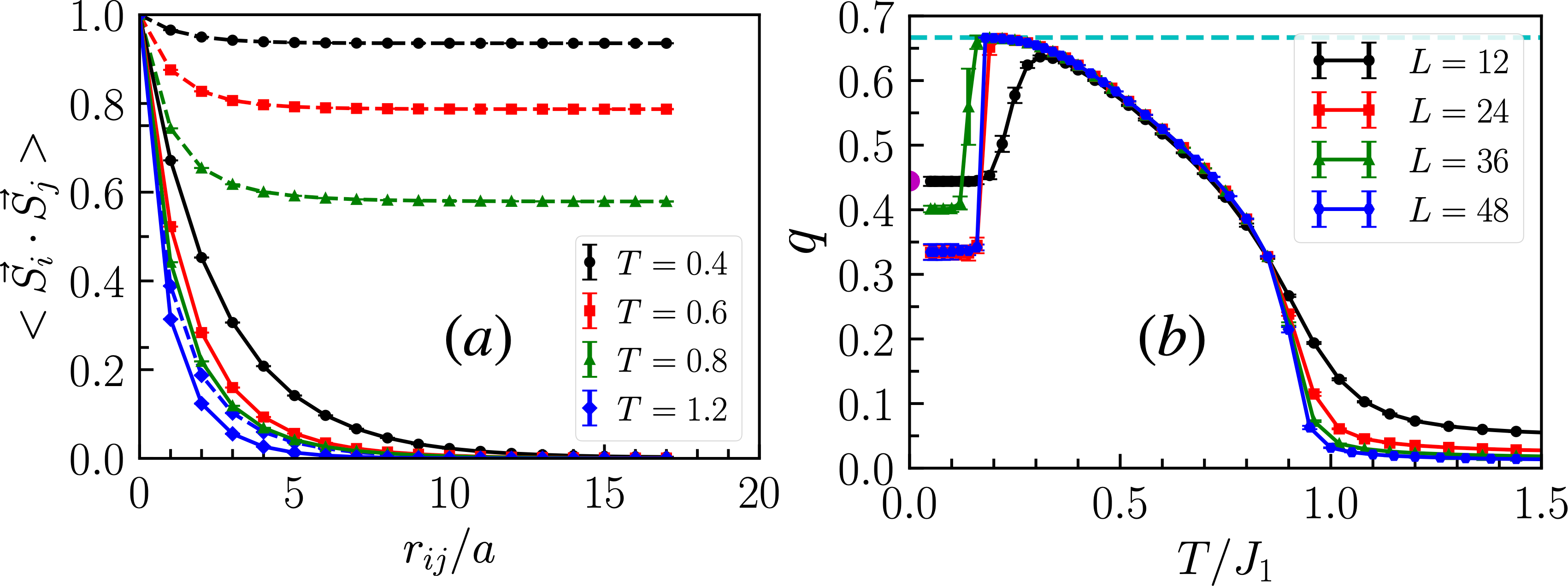}
\caption{\label{fig:corr_aniso_jdm}
MC results for the kagome ice model with spatially anisotropic $J_d$, using $J_{d}/J_{1}=0.4$ and $\Delta=0.2$.
(a) Spin-spin correlation function (in the global frame) for spins along the hexagon diagonals as a function of distance, with full (dashed) lines showing correlations along the weak (strong) bonds.
(b) Edwards-Anderson order parameter $q(T)$. The horizontal dashed line indicates $q=2/3$ and the red dot $q=4/9$.
}
\end{figure}

\paragraph{Strain pinning of partial order.---}
To corroborate the presence of partial order, we study the system in the presence of spatial anisotropy in $J_{d}$: For one of the three diagonal directions, we set $J_{d}\to J_{d}\left(1-\Delta\right)$. For $\Delta>0$, the partial order should be pinned along this direction because of the weaker couplings. Numerical results are shown in Fig.~\ref{fig:corr_aniso_jdm}.

The spin correlations in Fig.~\ref{fig:corr_aniso_jdm}(a) reflect order emerging along the strong bond direction but decaying correlations along the weak direction. We expect these spins to behave as ferromagnetic Ising chains, with a correlation length $\xi_{\rm chain}$ diverging as $T\to0$. This picture perfectly matches our results for the Edwards-Anderson order parameter, Fig.~\ref{fig:corr_aniso_jdm}(b). As long as $\xi_{{\rm chain}}$ is smaller than the system size $L$, we have $q\to2/3$, indicating that $1/3$ of the spins fluctuate. At low $T$ when $\xi_{{\rm chain}} > L$, there is a sudden drop in $q$, and it becomes equally probable for any of the three chains to fluctuate. For $L=12$, we recover the uniform result $q\to4/9$.  For larger system sizes,  however,  the runs do not reach equilibrium in this regime, as the magnetic states become markedly distinct above and below the drop in $q$,  rendering the parallel tempering algorithm inefficient. We have also computed $P\left(m_{x},m_{y}\right)$ for the case of anisotropic couplings, displaying only two peaks, corresponding to the selection of a single domain with the disordered chains running along the weaker bonds \cite{suppl}.

These results establish partial order in the kagome ice model with $J_{1,d}>0$. For $0<T<T_{c2}$, we see that $2/3$ of the spins show ferromagnetic long-range order, and the remaining $1/3$ form a set of effectively decoupled Ising chains which only order at $T=0$. The state below $T_{c2}$ thus spontaneously breaks a spatial $Z_3$ symmetry by selecting a chain direction.


\begin{figure}[!tb]
\includegraphics[width=1\columnwidth]{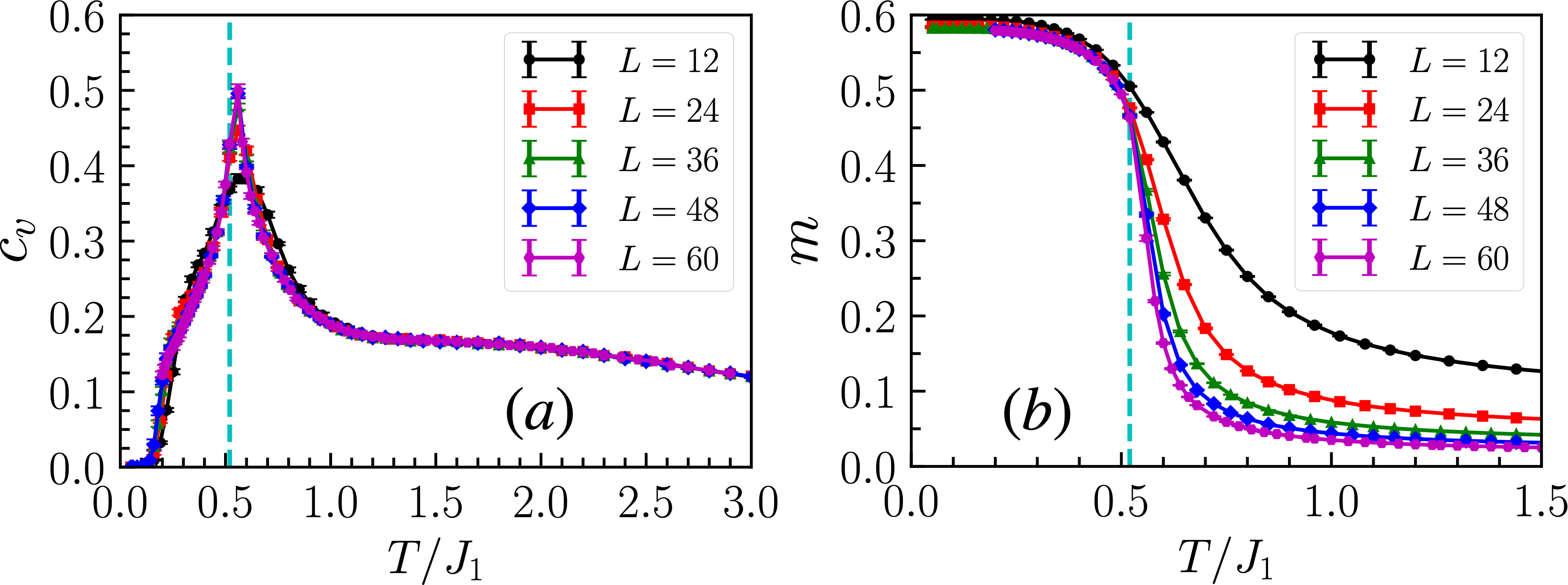}
\caption{\label{fig:cv_jdp}
MC results for the kagome ice model with antiferromagnetic $J_{d}/J_{1}=-0.4$.
(a) Specific heat $C(T)$ for different system sizes. The dashed line estimates $T_{c}$.
(b) Magnetization $m(T)$ \cite{suppl}.
}
\end{figure}

\begin{figure}[!bt]
\includegraphics[width=1\columnwidth]{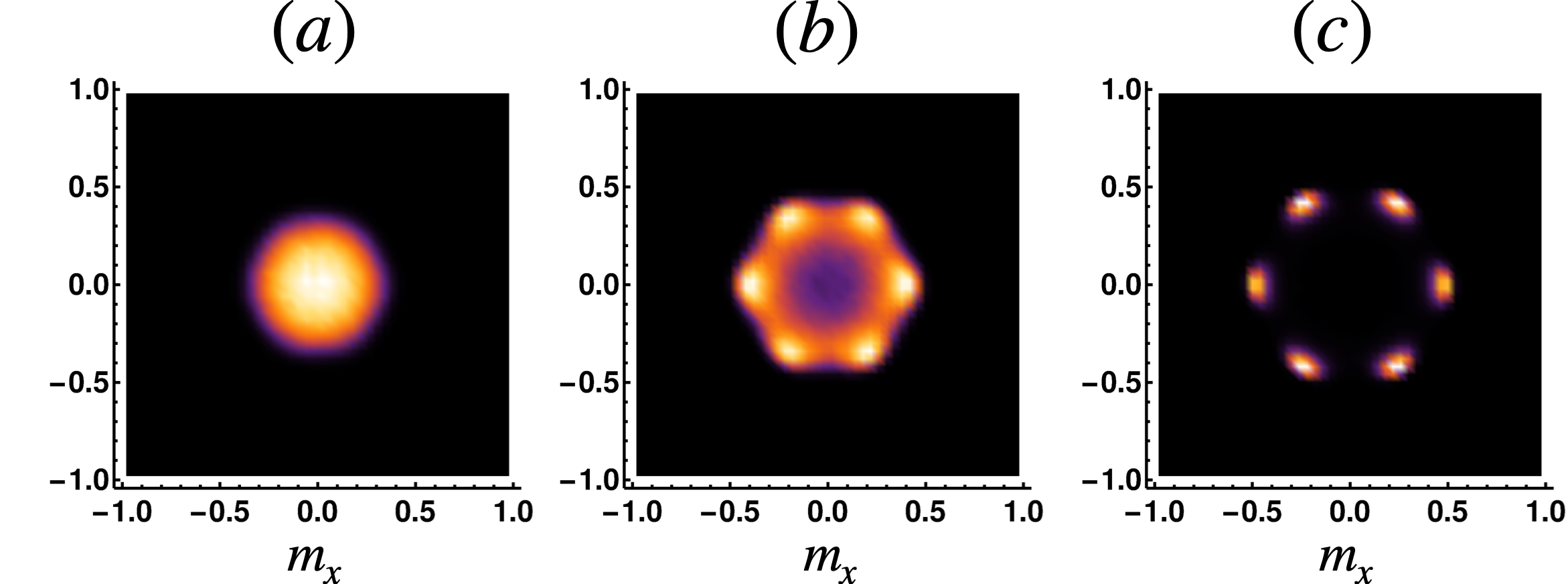}
\caption{\label{fig:P_mx_my_jdp}
Magnetization distribution $P(m_{x},m_{y})$ obtained for $L=24$ and $J_{2}/J_1=-0.4$.
(a) $T/J_1=0.6$,  (b) $T/J_1=0.56$, and (c) $T/J_1=0.52$.
The color scale is the same as in Fig.~\ref{fig:P_mx_my_jdm}.
}
\end{figure}

\paragraph{Partially ordered antiferromagnet.---}
Motivated by the results for $J_{d}>0$, we now study the model with $J_{d}<0$, searching for the partially ordered antiferromagnet in Fig.~\ref{fig:states}(d), which is likely of experimental relevance for {\fsso}.

As shown in Fig.~\ref{fig:cv_jdp}(a), the specific heat now shows a single peak whose temperature grows with $J_d$, consistent with a single phase transition. The magnetization, per up-triangle in the unit cell, approaches $1/\sqrt{3}$ as $T\to0$, Fig.~\ref{fig:cv_jdp}(b), as expected for partial order.



$P(m_{x},m_{y})$ displays the expected six-peak structure at low $T$,  Fig.~\ref{fig:P_mx_my_jdp}. We find no evidence for a ring-shaped $P\left(m_{x},m_{y}\right)$, consistent with the absence of a critical phase for all investigated values of $J_d$ \cite{suppl}. We believe this difference between the ferromagnetic and antiferromagnetic cases is rooted in the different magnetic unit cells, Figs.~\ref{fig:states}(c,d) and ~S1(a,b) \cite{suppl}.  For $J_d < 0$,  a given state and its time-reversal equivalent are related by a simple translation.  This suggests an emerging three-state clock anisotropy and a single transition \cite{jose1977}.


We have also studied the antiferromagnetic case in the presence of spatial anisotropy in $J_{d}$ to pin the disordered spins' manifold. Numerical results are shown in the supplement \cite{suppl}, again consistent with a partial order, with 1/3 of the spin forming antiferromagnetic Ising chains, which only order at $T=0$.


\begin{figure}[!tb]
\includegraphics[width=1\columnwidth]{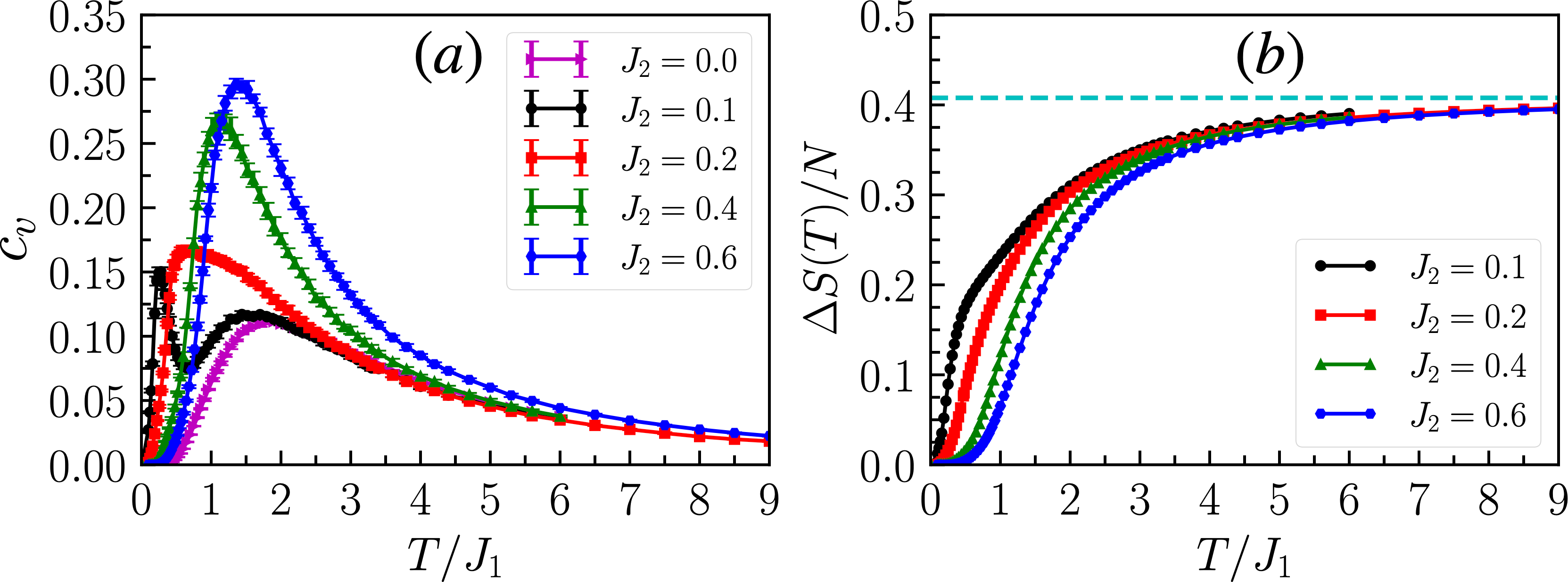}
\caption{\label{fig:cv_j2}
MC results for the kagome ice model with finite $J_2$, but $J_d=0$.
(a) Specific heat $C(T)$ and (b) entropy difference $\Delta S(T)=S(T)-S_0$ for different values of $J_2$ and system size $L=36$. The dashed line in (b) corresponds to $\ln 2- S_0$ with the exact value of $S_0$ from Ref.~\cite{colbois21}.
}
\end{figure}

\begin{figure}[!bt]
\includegraphics[width=1\columnwidth]{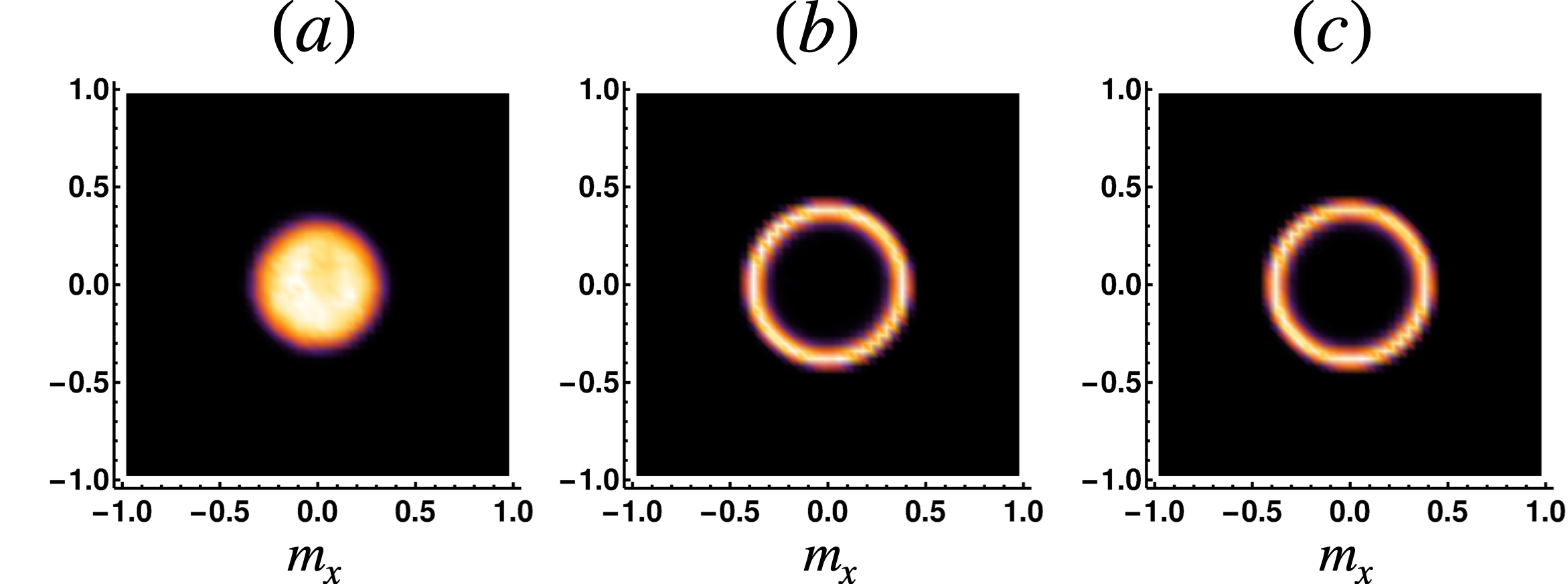}
\caption{\label{fig:P_mx_my_j2}
Magnetization distribution $P(m_{x},m_{y})$ obtained for $L=48$ and $J_{2}/J_1=0.1$.
(a) $T/J_1=0.2$,  (b) $T/J_1=0.1$, and (c) $T/J_1=0.05$.
The color scale is defined in Fig.~\ref{fig:P_mx_my_jdm}.
}
\end{figure}

\paragraph{No partial order driven by $J_2$.---}
We finally turn to the model initially proposed in Ref.~\onlinecite{wills02} to show partial order, namely the model with $J_{3,d}=0$ and ferromagnetic $J_2$ instead. Thermodynamic results are shown in Fig~\ref{fig:cv_j2}. For small $J_{2}$, we observe two broad peaks in $C(T)$, which tend to merge at larger $J_2$. Integrating $C(T)$, we find a finite residual entropy, consistent with this model displaying an extensive ground-state degeneracy \cite{takagi93,hamp18,colbois21}. A Pauling argument can estimate the entropy, considering that $J_1$ enforces ice rules on the original lattice whereas $J_2$ enforces ice rules on the respective kagome superlattices \cite{suppl}. A more precise value of the residual entropy is $S_{0}=0.2853N$ \citep{colbois21}, consistent with our data. This discussion implies that the two peaks in $C(T)$ correspond to crossovers where these ice rules are enforced by $J_1$ and $J_2$, respectively.

The magnetization data appear consistent with a KT transition in the limit $T\to0$, as proposed in Ref.~\onlinecite{takagi93}. For $T<J_1$, we observe a power-law behavior $m\sim L^{-\eta/2}$, with $\eta\to1/4$ as $T\to0$ \cite{suppl}. The absence of a finite-$T$ transition is consistent with the finite residual entropy we identify in Fig.~\ref{fig:cv_j2}(b). Another direct counter-evidence concerning partial disorder comes from the histogram $P(m_{x},m_{y})$ in Fig.~\ref{fig:P_mx_my_j2}. Here, we see a continuous ring structure down to the lowest $T$, akin to the critical phase discussed above, with no sign of the six-peak structure.


We conclude that the model with finite $J_2$ (and $J_{3,d}=0$) does not realize a state with partial order. Instead, its low-$T$ phase is critical with power-law correlations, and the partially ordered state represents only one of the many possible low-$T$ states of the model.


\paragraph{Summary.---}
For variants of kagome spin ice, with first-neighbor and diagonal third-neighbor couplings, we have established the existence of partially ordered magnetic states where $1/3$ of the spins continue to fluctuate down to $T\to0$ in the background of either ferromagnetic or antiferromagnetic bulk order. The origin is intense frustration and dimensional reduction: The fluctuating spins form effective Ising chains, which experience no mean field and are entropically disordered at $T>0$.

This peculiar state is robust with respect to additional (weak) second-neighbor coupling $J_2$ and a third-neighbor coupling $J_3$ across a site,  Fig~\ref{fig:states}(a).  Both perturbations cause no mean field on the disordered sites, thus preserving the partially disordered state.  However,  $J_3$  locks neighboring disordered chains into a long-range ordered state,  resulting in a weakly ordered state at low $T$ \cite{suppl}.

Our findings are likely relevant to {\fsso}, and we propose to combine specific heat measurements with local probes to disentangle the energy scales at which the ice rules are imposed and the partial order emerges. Moreover, measurements under uniaxial strain, as simulated here, will be key to nail down the one dimensional nature of the partially ordered state.
Extending the present theory beyond the limit of infinite Ising anisotropy and in the presence of a magnetic field is the subject of ongoing work.


\acknowledgments

We thank H.-H. Klauss and R. Sarkar for discussions and collaborations on related topics.
We acknowledge support by the DFG through SFB 1143 (project id 247310070) and the W\"urzburg-Dresden Cluster of Excellence on Complexity and Topology in Quantum Matter---\textit{ct.qmat} (EXC 2147, project id 390858490).
ECA was supported by CNPq (Brazil), Grant No. 302823/2022-0, and FAPESP (Brazil), Grant No. 2021/06629-4.  ECA also acknowledges the hospitality of TU Dresden, where part of this work was performed.


\bibliography{ice}

\end{document}


\title{
Supplemental material for: \\
Partial magnetic order in kagome spin ice
}

\author{Eric C. Andrade}
\affiliation{Instituto de F\'isica, Universidade de S\~ao Paulo, C.P. 66318,
05315-970, S\~ao Paulo, SP, Brazil}
\author{Matthias Vojta}
\affiliation{Institut f\"ur Theoretische Physik and W\"urzburg-Dresden Cluster
of Excellence ct.qmat, Technische Universit\"at Dresden, 01062 Dresden,
Germany}

\date{\today}

\maketitle


\section{Characterization of partial order}

\begin{figure}[tb]
\includegraphics[width=1\columnwidth]{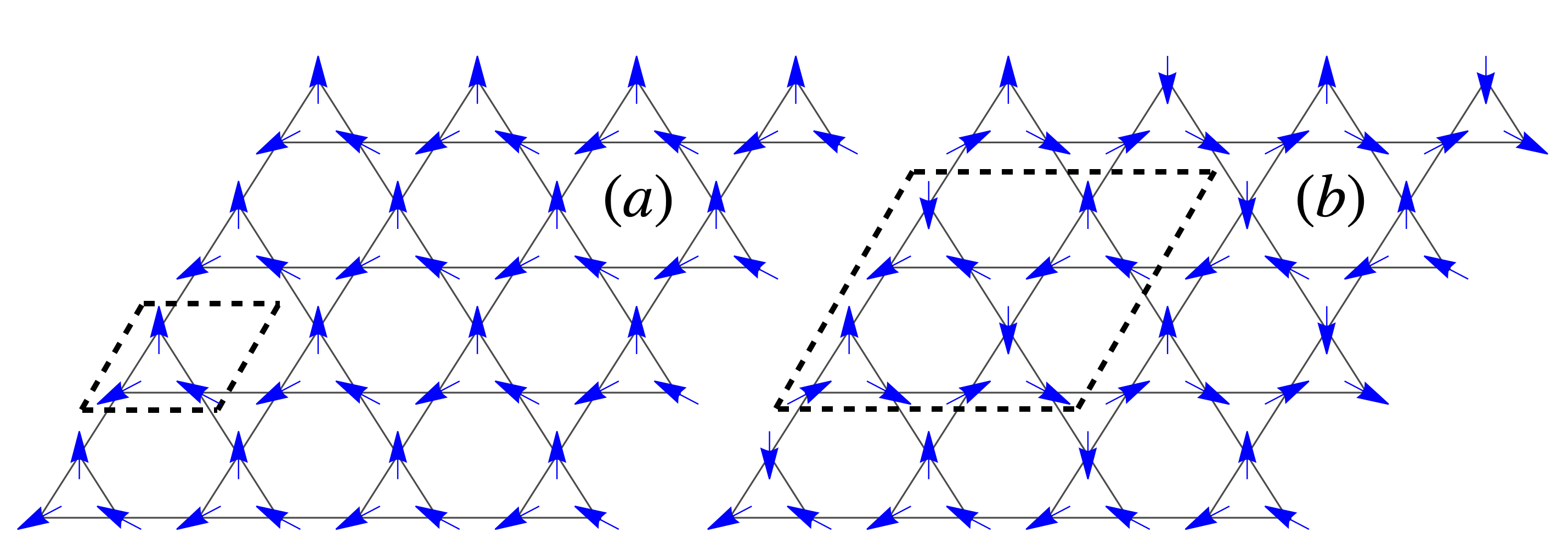}
\caption{\label{fig:configMC}
Snapshots of a low-$T$ MC spin configuration in the global frame for (a) $J_d>0$ and (b) $J_d<0$. The dashed lines indicate the magnetic unit cell in each case.
}
\end{figure}

Representative snapshots of low-$T$ spin configurations obtained from our Monte-Carlo (MC) simulations of the model with finite $J_1$ and $J_d$ are shown in Fig.~\ref{fig:configMC}. The partially ordered states emerge as several such configurations are averaged during the MC simulation.
To characterize these partially ordered phases and differentiate them from putative fully ordered phases, we use our MC data to compute the magnetization $\boldsymbol{m}$ in the global frame (for each up-triangle in the magnetic unit cell), the static spin structure factor $S(\boldsymbol{k})$, as well as the Edward-Anderson order parameter $q$, a quantity designed to capture the spin-freezing independent of long-range order \cite{lee07}.

First, we discuss the magnetization in the global frame, illustrated in Fig.~1(b) of the main text. We recall that $\boldsymbol{S}_{i}=\sigma_{i}\hat{\boldsymbol{e}}_{i}$, with the local axes given by $\hat{e}_\alpha=\left(\cos\theta_\alpha, \sin\theta_\alpha\right)$, with $\alpha=A,B,C$ the sublattice index and $\theta_A=7\pi/6$,  $\theta_B=11\pi/6$,  $\theta_C=\pi/2$. We compute $\boldsymbol{m} = (3N_t)^{-1}\sum_t \sum_{i\in t} \boldsymbol{S}_i$ where $\sum_t$ runs over all up-triangles (every fourth up-triangle) in the ferromagnetic (antiferromagnetic) case, respectively, and $N_t$ counts these up-triangles.
Consider now the partially ordered configuration where the disordered spins occupy the sublattice $C$, $\langle\sigma_C\rangle=0$ where $\langle\ldots\rangle$ denotes MC (equivalently, thermodynamic) average. The magnetization of this partially ordered state is $\langle\boldsymbol{m}\rangle=\left(-\sqrt{3},0\right)/3$ and thus $m\equiv|\langle\boldsymbol{m}\rangle|\to1/\sqrt{3}$ as $T\to0$. The other local configurations in Fig.~1(b) produce the same result for $m$. In contrast,  $\boldsymbol{m}=\left(-\sqrt{3},1\right)/3$,  in a fully ordered state with the same unit cell,  Fig.~\ref{fig:configMC}(a),  and $m \to 2/3$ as $T \to 0$. Hence, this observable serves as an indicator of partial order.

We also compute the histogram of $\boldsymbol{m}=\left(m_{x},m_{y}\right)$ as defined above, i.e., the average is taken over all up-triangles (every fourth up-triangle) in the ferromagnetic (antiferromagnetic) case. The histogram is generated along the MC run from the different MC configurations. We can deduce the expected values from Fig.~1(b) of the main text. Taking again the partially ordered configuration with disordered spins occupying sublattice $C$ as an example, we have that $\boldsymbol{m}=\left(-\sqrt{3},0\right)/3$ -- note that $\langle\sigma_C\rangle=0$ arises now from the proper average over all the sample -- and hence $\theta=\pi$ in $\tan\theta=m_{y}/m_{x}$. The partially ordered configuration with disordered spins occupying sublattice $B$ yields $\boldsymbol{m}=\left(1/\sqrt{3},1\right)/2$ and thus $\theta=\pi/3$, etc. Therefore, for a partially ordered state, we expect $\theta=n\pi/3$, $n=0,1,\ldots,5$, arising from the three possible sublattice choices and their time reversal. In contrast, the ordered configuration in Fig.~\ref{fig:configMC}(a) gives $\boldsymbol{m}=\left(-\sqrt{3},1\right)/3$, leading to $\theta=5\pi/6$.  Therefore,  configurations with three spins per triangle satisfying the ice rules produce $\theta=(2n+1)\pi/6$, $n=0,1,\ldots,5$. The histogram of $P(m_{x},m_{y})$  differentiates both cases and thus provides a strong signature of partially ordered states.

We further characterize the spin state via the static spin structure factor $\mathcal{S}\left(\boldsymbol{k}\right)=
(1/N) \sum_{ij} \langle \boldsymbol{S}_i \cdot \boldsymbol{S}_j \rangle e^{-i(\boldsymbol{r}_{i}-\boldsymbol{r}_{j})\cdot\boldsymbol{k}} = \langle\boldsymbol{S}\left(\boldsymbol{k}\right)\cdot\boldsymbol{S}\left(-\boldsymbol{k}\right)\rangle$,  where $\boldsymbol{S}\left(\boldsymbol{k}\right)=N^{-1/2}\sum_{j}e^{-i\boldsymbol{r}_{j}\cdot\boldsymbol{k}}\boldsymbol{S}_{j}$ is the Fourier transform of a given spin configuration.  This quantity is unable to differentiate ordered from partially ordered states.  However,  it can distinguish states with different magnetic unit cells, thus probing the cases of positive and negative $J_{d}$. Moreover,  sharp Bragg peaks in $\mathcal{S}\left(\boldsymbol{k}\right)$ confirm that the disordered sites occupy well-defined positions.

Finally,  we consider the Edwards-Anderson order parameter, as defined in the main text, measuring the local overlap of two independent spin configurations (replicas) at a given MC step.  In a given partially ordered state,  as the one in Fig.~1(b) of the main text,  we have $|q|=2/3$ because $1/3$ of the spins remain fluctuating in this state. However, the MC simulation equally samples all possible spatial realizations of the partial order, and we obtain $|q|=1/3$ when computing the overlap between different partial order configurations.  We thus get $|q|=4/9$, on average. This is the value quoted in the main text. In contrast, a fully ordered state with a unique ordering gives $|q|=1$.


\section{Additional data and discussion for the $J_d$ model}

Here, we show and discuss additional MC results obtained for the kagome ice model with finite $J_d$ and $J_{2,3}=0$, first for the model with three equivalent bond directions and then for the modified (strained) model where the $J_d$ bonds along one direction are weakened.

\begin{figure}[tb]
\includegraphics[width=1\columnwidth]{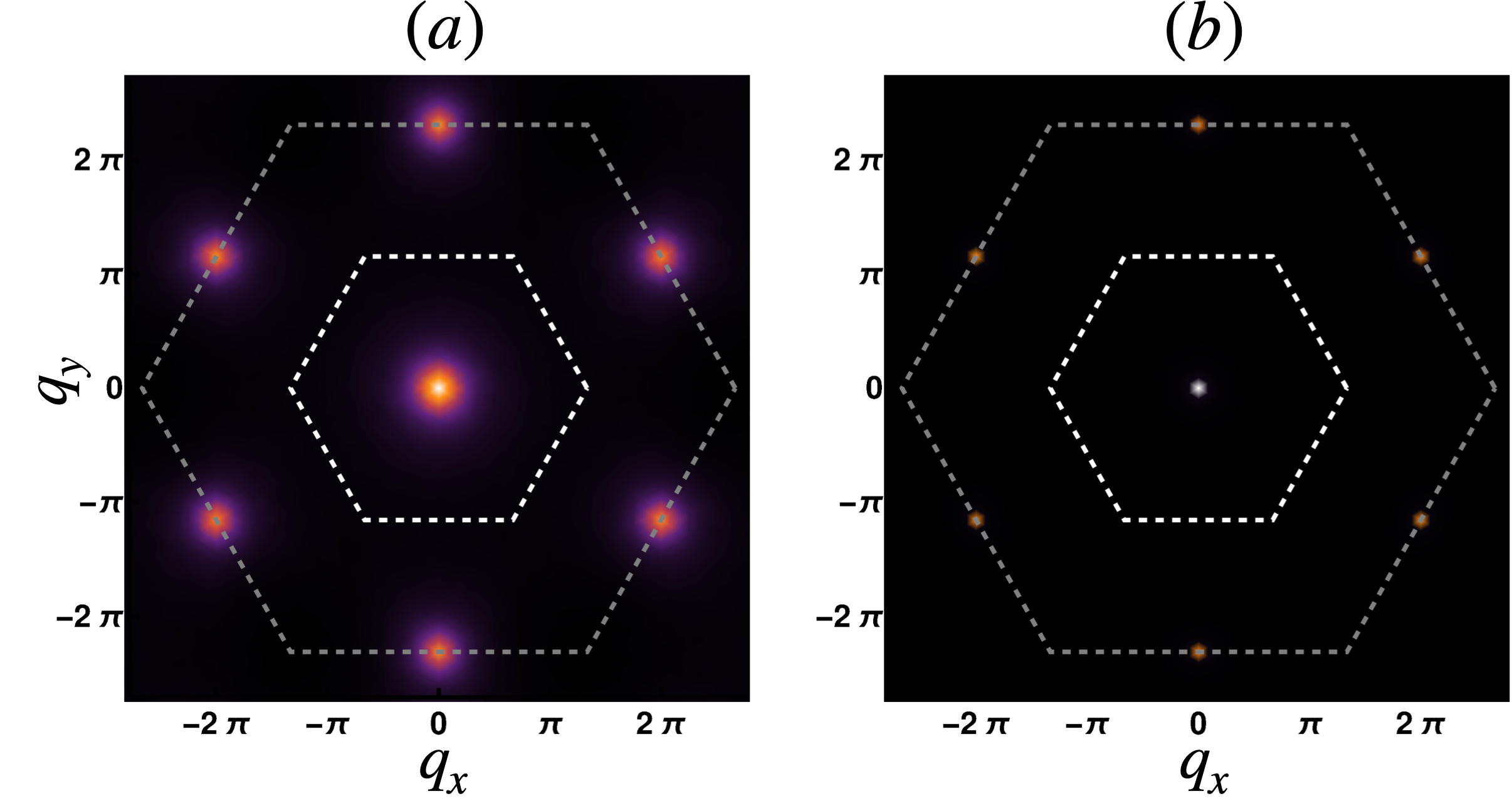}
\caption{\label{fig:sk_fm}
Static structure factor for $J_{d}/J_{1}=0.4$.  (a) $T/J_{1}=1.2$ and (b) $T/J_{1}=0.6$.  The dashed lines correspond to the first and second Brillouin zones,  respectively.  The color scale is the same as in Fig.~3 of the main text.}
\end{figure}

\begin{figure}[tb]
\includegraphics[width=1\columnwidth]{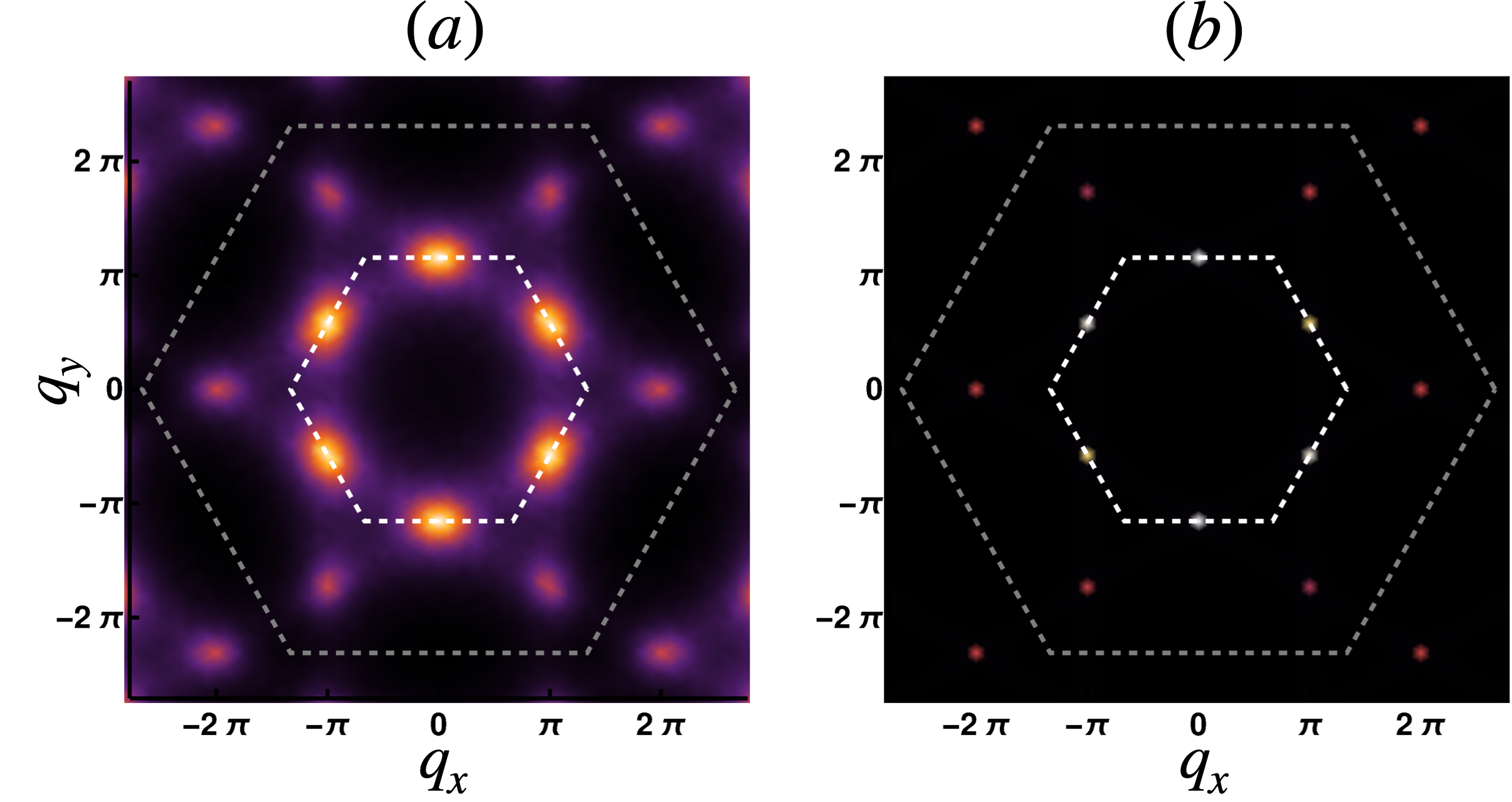}
\caption{\label{fig:sk_afm}
Same as Fig.~\ref{fig:sk_fm}, but now $J_{d}/J_{1}=-0.4$ and (a) $T/J_{1}=0.8$ and (b) $T/J_{1}=0.4$.
}
\end{figure}

\subsection{Isotropic model}

In Fig.~\ref{fig:sk_fm}, we show the static structure factor for the partially ordered state appearing for $J_{d} >0$.  Because its magnetic unit cell is the same as the original one, we observe the Bragg peaks at those positions expected for a regular ferromagnetic state \cite{messio11}.  Because the disordered sites appear periodically, all Bragg peaks are sharp, and this quantity has no immediate signature of partial order.

In Fig.~\ref{fig:sk_afm}, we display the same quantity now for $J_{d} <0$. Here, the magnetic unit cell corresponds to one of the octahedral states found in the Heisenberg model in the kagome lattice \cite{messio11,oliviero22},  and we thus expect the Bragg peaks at the corresponding positions, in agreement with our MC results. Again, the Bragg peaks are sharp due to the periodic nature of the disordered sites.

\begin{figure}[tb]
\includegraphics[width=1\columnwidth]{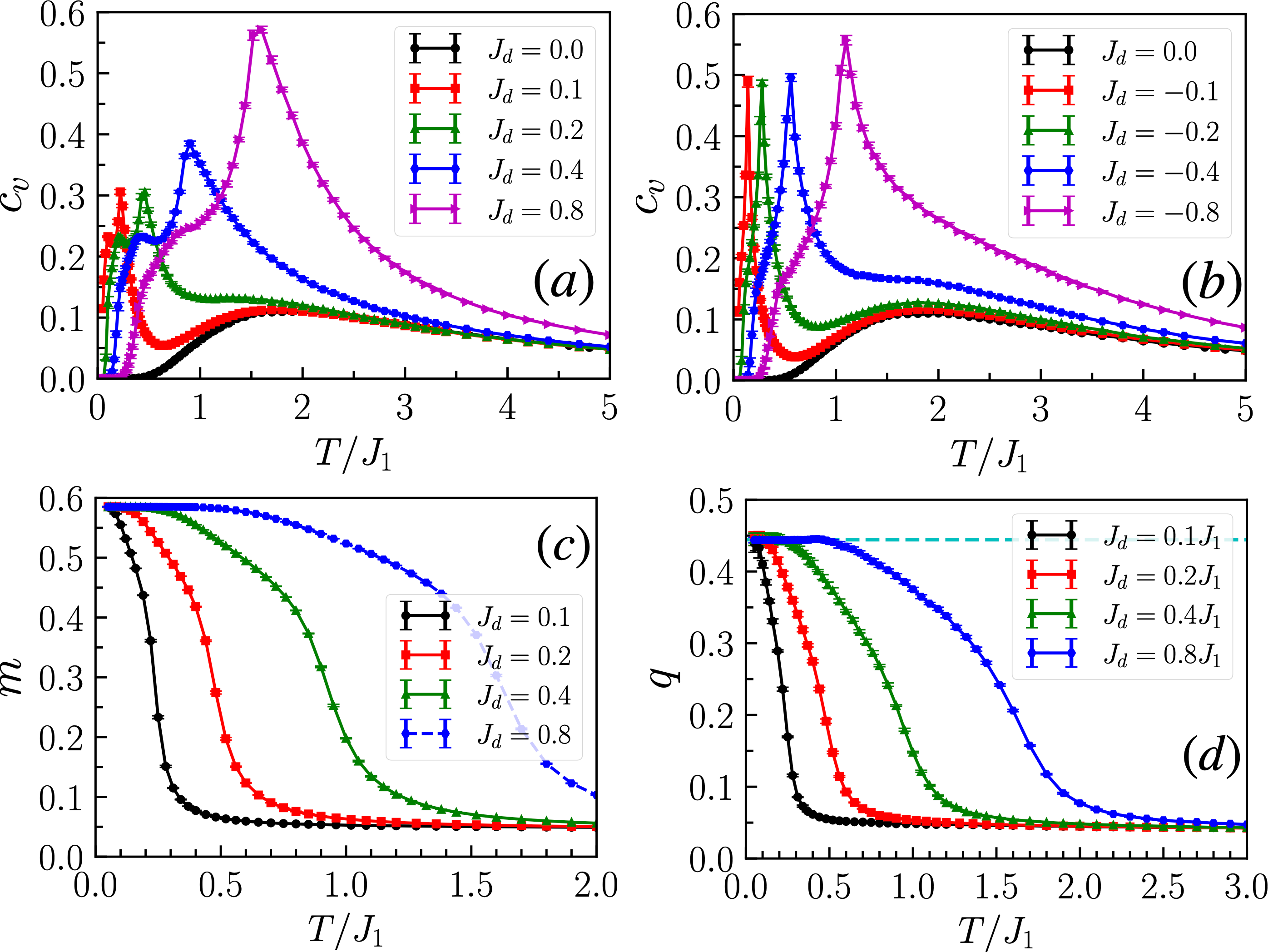}
\caption{\label{fig:extra_jd_iso}
Specific heat as function of temperature for $L=48$ and (a) $J_{d}>0$ and (b) $J_{d}<0$.  (c) Magnetization as a function of temperature for $L=24$ and $J_{d}<0$.  The dot at $T=0$ marks the value $m=1/\sqrt{3}$.  (d) Edwards-Anderson order parameter as a function of the temperature for $L=12$ and $J_{d} > 0$.  The dashed line indicates the value $q=4/9$.}
\end{figure}

We now present additional thermodynamic results for different values of $J_d$ in Fig.~\ref{fig:extra_jd_iso}.  The evolution of the specific-heat curves with $J_d$, Fig.~\ref{fig:extra_jd_iso}(a,b), indicates that the transition temperatures, as given roughly by the peak position in $C(T)$, scale with $J_d$.  There is also an ice-rule bump around $T=1.8J_1$, which does not move much as a function of $J_d$ but merges with the transition peaks as $J_d$ increases and becomes invisible. From this data, the main difference between positive and negative $J_d$ is the presence of a second finite-$T$ transition for $J_d > 0$.

Fig.~\ref{fig:extra_jd_iso}(c) shows the absolute value of the magnetization for $J_d < 0$.  We expect $m\to1/\sqrt{3}$ as $T\to0$ in the partially ordered state, and we recover this result for all values of $J_d$ we study. Fig.~\ref{fig:extra_jd_iso}(d) shows the Edwards-Anderson order parameter for $J_d > 0$ as it goes to the expected $4/9$ as we enter the partially ordered state.  Because it is hard to equilibrate the MC simulation at low $T$,  we consistently obtain $|q| \to 4/9$ for small system sizes only.

\begin{figure}[b]
\includegraphics[width=1\columnwidth]{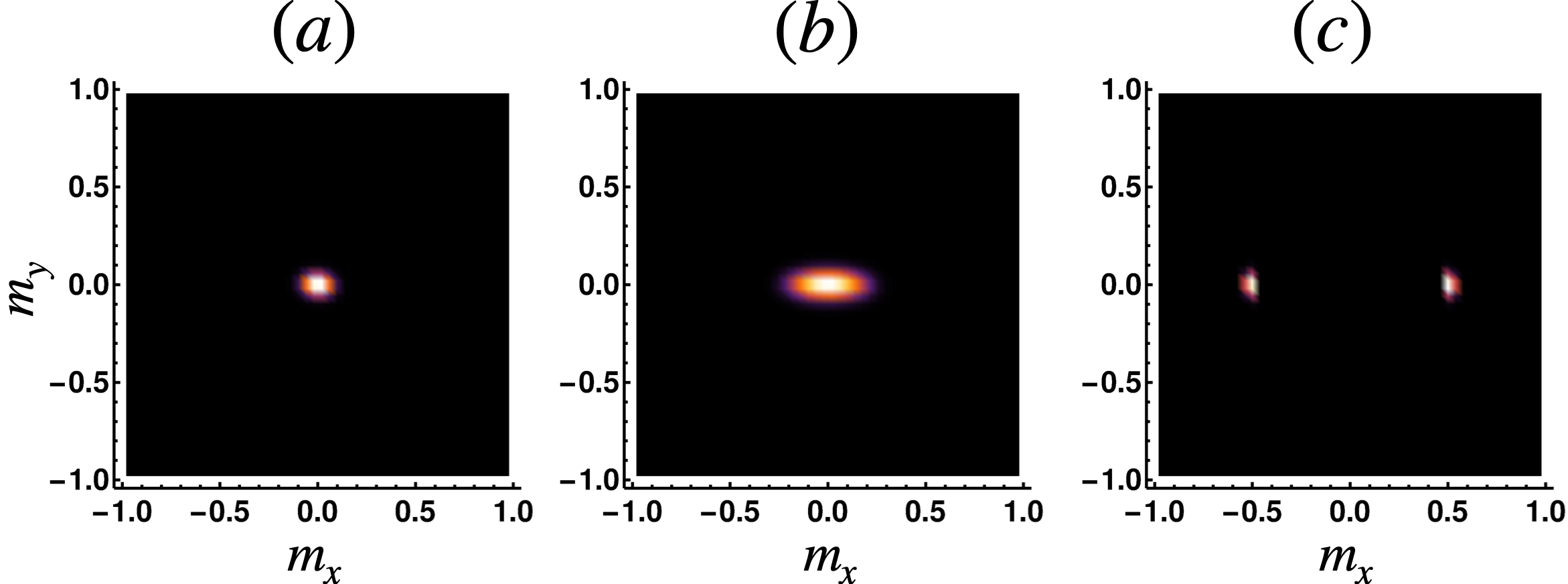}
\caption{\label{fig:P_mx_my_jdm_aniso}
Distribution $P\left(m_{x},m_{y}\right)$ obtained for $L=48$, $J_{d}/J_{1}=0.4$, and $\Delta=0.2$.
(a) $T/J_{1}=1.2$,  (b) $T/J_{1}=1.0$,  and (c) $T/J_{1}=0.6$. The color scale is the same as in Fig.~3 of the main text.}
\end{figure}

\subsection{Anisotropic model}

For the case of spatially anisotropic couplings and $J_{d} >0$, we present the histogram of the magnetization in Fig.~\ref{fig:P_mx_my_jdm_aniso}. Due to the anisotropy, we pin the disordered sites to a particular direction, and thus, we observe only two peaks in $P\left(m_{x},m_{y}\right)$.  The anisotropy is present at intermediate temperatures as the signal around the origin becomes an ellipse, and the intermediate critical phase is absent since the real-space directions become inequivalent.  This scenario is consistent with the results presented in the main text.

\begin{figure}[bt]
\includegraphics[width=1\columnwidth]{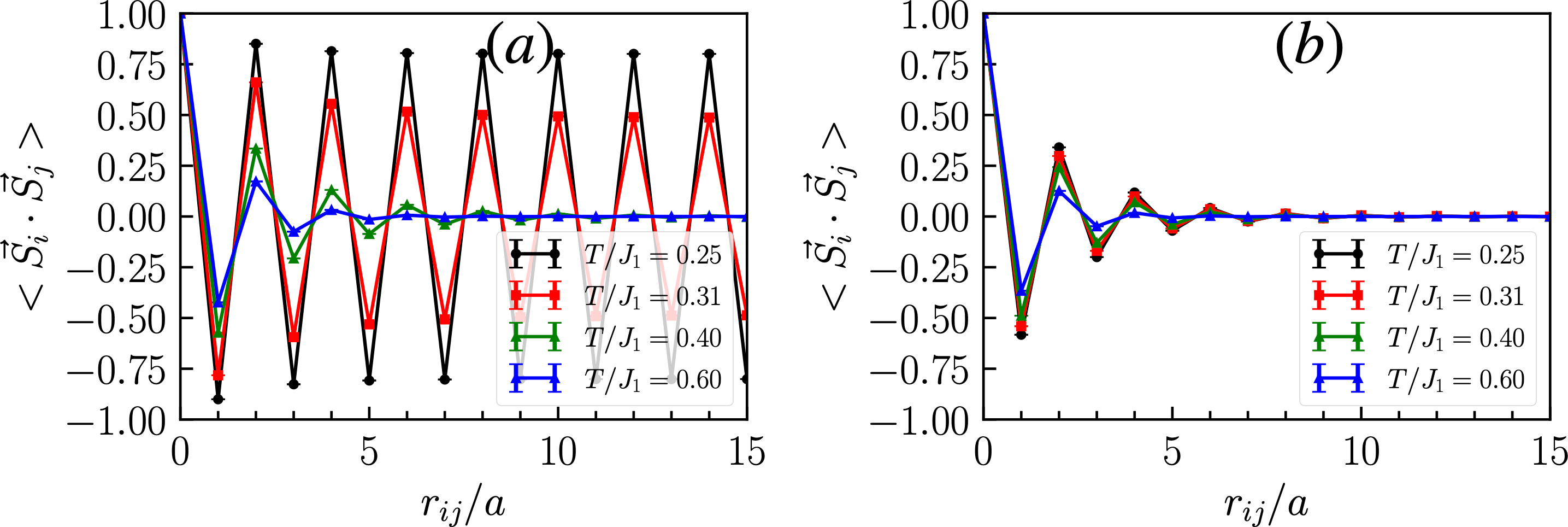}
\caption{\label{fig:corr_aniso_jdp}
Spin-spin correlation function for spins (in the global frame) along the diagonals of the hexagon as a function of distance between spins, shown along the directions of (a) strong bonds and (b) weak bonds.  These results are for $J_{d}/J_{1}=-0.4$, anisotropy parameter $\Delta=0.2$, and $L=36$.
}
\end{figure}

\begin{figure}[tb]
\includegraphics[width=1\columnwidth]{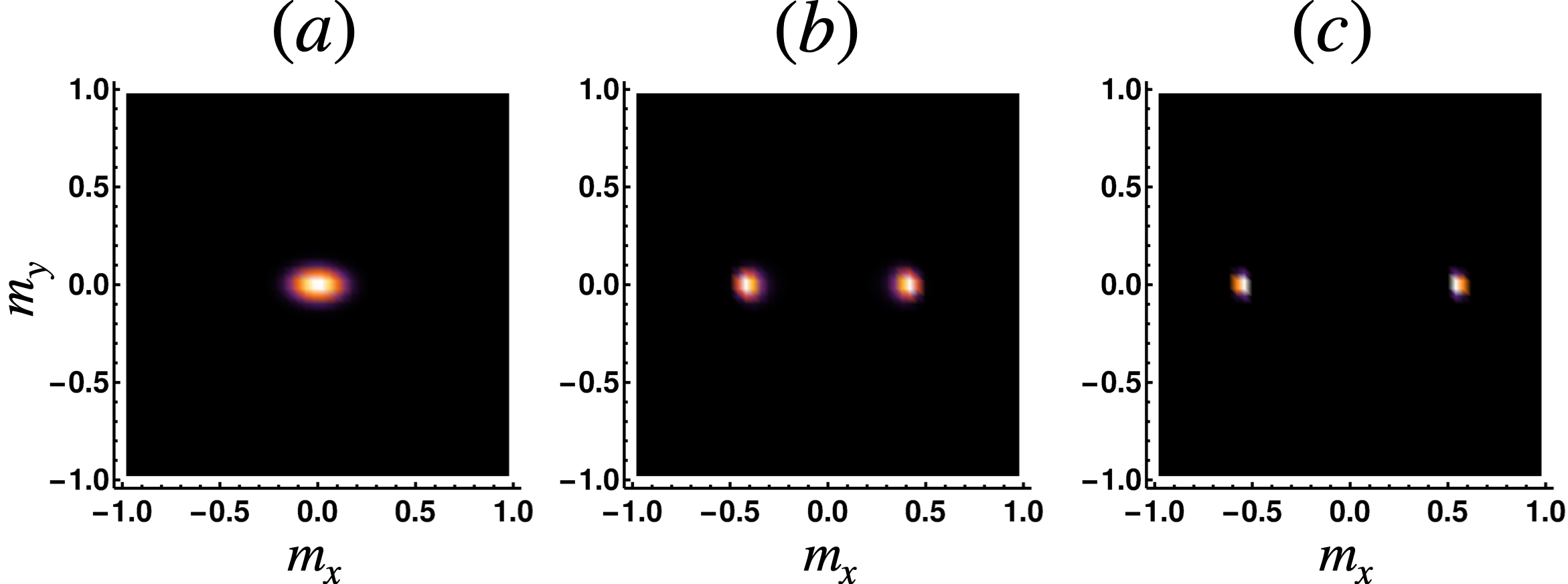}
\caption{\label{fig:P_mx_my_jdp_aniso}
Magnetization distribution $P\left(m_{x},m_{y}\right)$ obtained for $L=48$, $J_{d}/J_1=-0.4$, and $\Delta=0.2$.
(a) $T/J_{1}=0.72$,  (b) $T/J_{1}=0.5$,  and (c) $T/J_{1}=0.4$.
}
\end{figure}

Now, for $J_{d} <0$  and spatially anisotropic couplings, we first show the spin-spin correlation function along the diagonals of the hexagon in Fig.~\ref{fig:corr_aniso_jdp}.  The antiferromagnetic character of the correlations becomes evident as the correlation alternates signs.  The critical temperature in the isotropic model is around $T_{c} \sim J_{1}/2$,  and the development of long-range order is visible along the strong bonds below this value.  However,  the one-dimensional chains do not order along the weaker bonds,  consistent with the partially disordered sites pinning along them.  The histograms of the magnetization in Fig.~\ref{fig:P_mx_my_jdp_aniso} are compatible with this scenario as we only observe two peaks in $P\left(m_{x},m_{y}\right)$,  similar to Fig.~\ref{fig:P_mx_my_jdm_aniso}.

\section{Additional data and discussion for the $J_2$ model}

We present the Pauling estimate for the ground-state degeneracy of this model. The $J_{2}$ interactions form three independent kagome superlattices on which they try to enforce the ice rules. Each of these superlattice has $N_{{\rm triang}}^{\prime}$ = $N_{{\rm triang}}/3$ triangles, with $N_{{\rm triang}}=2N/3$ the number of triangles in the original kagome lattice. The number of possible states is then
\begin{align}
\Omega
&=2^{N}\times\left(\frac{6}{8}\right)^{N_{{\rm triang}}}\times\left(\frac{6}{8}\right)^{3N_{{\rm triang}}^{\prime}} \notag\\
&=2^{N}\times\left(\frac{6}{8}\right)^{2N_{{\rm triang}}}\simeq\left(1.363\right)^{N}.
\label{eq:entropy_J2}
\end{align}
The first term above imposes the ice rules due to $J_{1}$, whereas the second one imposes the one due to $J_{2}$. The residual entropy is $S_{0}=\ln\Omega=0.309N$.

In the MC simulations, we observe two peaks in the specific heat for small $J_{2}$. The one at high temperature comes from $J_{1}$ imposing the ice rules, and the lower peak comes from $J_{2}$ selecting the subset of states which set the ice rules in the superlattice
\citep{wills02,colbois21}. The two peaks merge for larger $J_{2}$. The specific heat and the entropy difference, $\Delta S\left(T\right)=\int_{0}^{T}dT^{\prime}\,c_{v}\left(T^{\prime}\right)/T^{\prime}$,
are shown in Fig.~7 of the main paper, and are consistent with the absence of long-range order, displaying virtually no size dependence and a residual entropy which is close to the exact value of $S_{0}=0.2853N$ \citep{colbois21}.

\begin{figure}[tb]
\includegraphics[width=1\columnwidth]{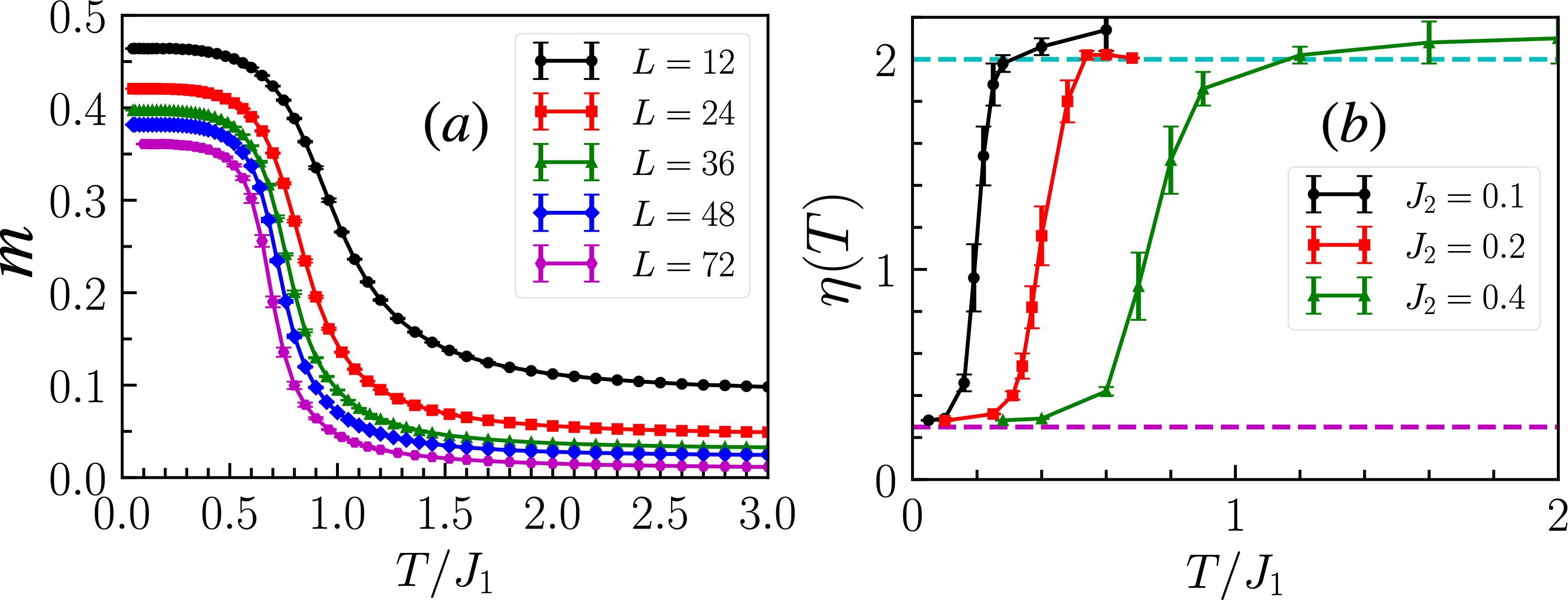}
\caption{\label{fig:eta_j2}
(a) Magnetization $m(T)$ for the $J_2$ model for different system sizes and $J_{2}/J_1=0.40$. 
(b) Effective power-law exponent $\eta$ as function of temperature. The point where it deviates from $2$ roughly coincides
with the low$-T$ peak in $C(T)$, indicating that $J_{2}$ is enforcing the ice rules.
}
\end{figure}

The nature of the ground state was discussed in Ref. \citep{takagi93}.  There, the authors propose a KT transition as $T\to0$.  Our results are consistent with theirs if we consider the magnetization in the global frame as the order parameter. At the mean-field ordering temperature of order $2J_{2}$, we have $\left|\boldsymbol{m}\right|>0$, whereas its phase is still disordered. We estimate the exponent $\eta$ using $\left|\boldsymbol{m}\right|\sim L^{-\eta/2}$,  since we observe that the absolute value of the magnetization decreases with the system size,  Fig.~\ref{fig:eta_j2}.  Our results show that $\eta\to1/4$ as $T\to0$ and that $m$ never reaches the value of $1/\sqrt{3}$ expected for the partially disordered state,  indicating that this state is one of the many contributing to the extensive residual entropy in this parameter range.

\section{Stability of the partially ordered state}

\begin{figure}[b]
\includegraphics[width=1\columnwidth]{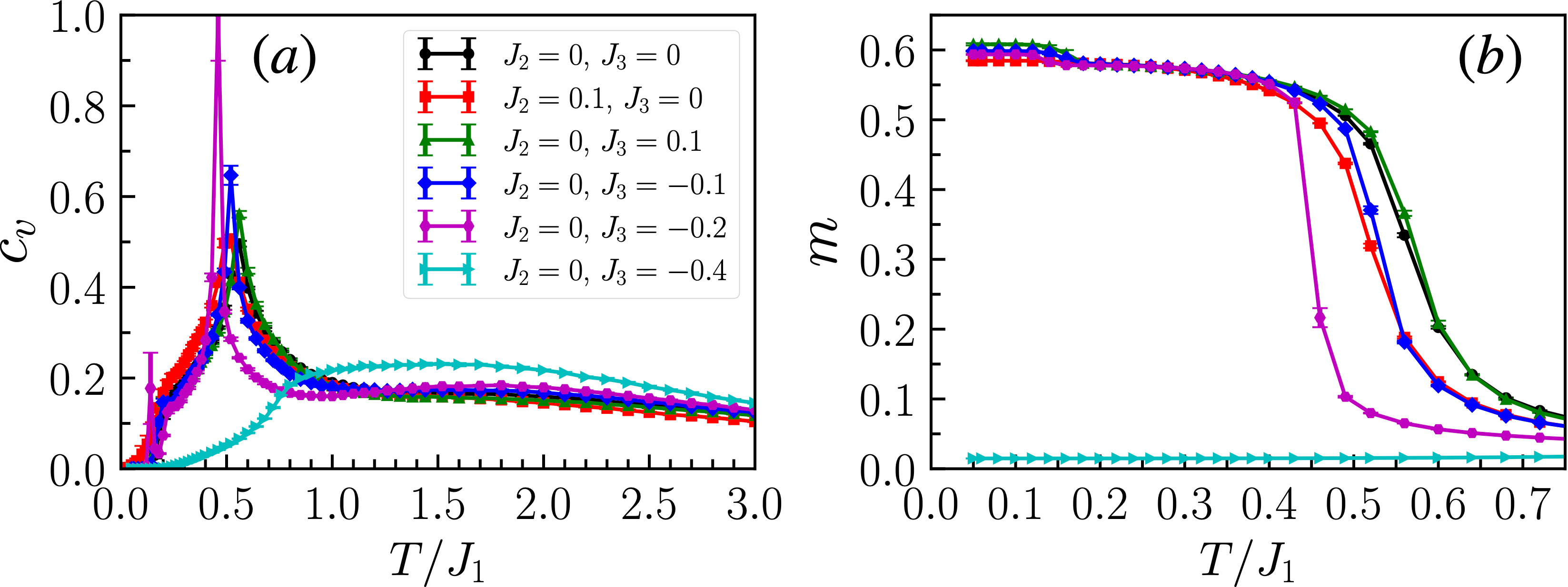}
\caption{\label{fig:mag_j2_j3}
Stability of the partially ordered phase for $J_{d}/J_{1}=-0.4$ with respect to different values of the couplings $J_2$ and $J_3$ in Eq.  (1) of the main text. 
(a) Specific heat as a function of the temperature.
(b) Magnetization as a function of temperature.  The color coding is the same as in (a).
All results are for $L=48$.
}
\end{figure}

The primary purpose of our numerical study was to show that $J_d$ stabilizes partial order,  in contrast to earlier claims that $J_2$ could be responsible for generating this state \citep{wills02}.  It is then natural to ask whether this partial order exists beyond this specific model,  and our data show that, remarkably,   additional (weak) second-neighbor coupling $J_2$ and a third-neighbor coupling across a site $J_3$ leave the partial order intact.  The reason is that both couplings generate a zero exchange field for the partially ordered state.  However, $J_3$ has a non-trivial effect because it couples neighboring disordered chains,  and long-range order emerges at low temperatures.  

Fig.  \ref{fig:mag_j2_j3} shows the specific heat and the magnetization as a function of the temperature for a fixed system size,  $L=48$,  and $J_d=-0.4J_{1}$.  For $ 0 < J_2 \lesssim 0.3J_{1}$,  we see no extra features on $c_{v}$ and $m \to 1/\sqrt{3}$ as $T\to 0$,  confirming the partial order remains stable.  For $J_{2} = 0.4 J_{1}$, we find a finite residual entropy,  indicating a disordered ground state.  For $J_2=-0.1J_{1}$, the partial order survives (not shown).  For larger negative values of $J_2$ we expect the $\sqrt{3} \times \sqrt{3}$ state to appear~\cite{chern11,chern12,colbois22}.  

For a finite $J_3$,  there is a second feature in $c_v$ for $T \approx 0.1J_1$,  which is more evident for $J_{3}/J_{1}=-0.2$.  We associate this point with the coupling between the disordered chains induced by $J_3$ and the development of long-range order.  This scenario is supported by the magnetization curve,  as $m$ deviates from the $1/\sqrt{3}$ value at low $T$.  For the ordered state in Fig.  \ref{fig:configMC}(b), we expect $m \to 2/3$ as $T \to 0$.  One would need considerably larger system sizes to reach this value because the correlation length is exceedingly large in this low $T$ regime.  If $|J_{3}| \ge |J_{d}|$,  the partial order state is lost, and we move into a disordered ground state \citep{colbois22}.  This is confirmed by the featureless specific heat curve and the absence of magnetization for all $T$.  We find a residual entropy entropy of $S_0=0.28N$ for $J_{3}/J_{1}=-0.4$ and $-0.6$,  a value consistent with $S_0/N=0.26413(5)$ observed for $-0.5 < J_3=J_d < 0$ in Ref.~\cite{colbois22}.


\bibliography{ice}